\documentclass[review,preprint]{revtex4}
\usepackage[dvips]{graphicx}
\usepackage{t1enc,amsmath}
\usepackage{amsfonts}
\usepackage{amssymb}
\begin{document}
 
\title{Half-wave plate based on a birefringent metamaterial in the visible range}
\author{T Zeghdoudi$^{1,2*}$, Z Kebci$^{1}$, A Mezeghrane$^{1}$, A Belkhir$^1$ and F I Baida$^{2}$}
\affiliation{$^{1}$ Laboratoire de Physique et Chimie Quantique, Universit\'{e} Mouloud Mammeri, Tizi-Ouzou, Algeria\\
$^{2}$ Institut FEMTO-ST, UMR 6174 CNRS, D\'epartement d'Optique P. M. Duffieux,
 Universit\'{e} Bourgogne Franche--Comt\'{e}, 25030 Besan\c{c}on Cedex, France
$^*$email: thinhinane.zeghdoudi@ummto.dz}
\begin{abstract}
In the present paper, a half-wave plate (HWP) based on a birefringent metamaterial is numerically designed to operate in the visible range. The proposed structure consists of an array of double-pattern perpendicular rectangular aperture (RAA) engraved into opaque silver film deposited on a glass substrate. One of the apertures is glass filled. The operating principle of this plate is based on the excitation and the propagation of one guided mode inside each aperture but with different effective index. At the output side, a phase difference is obtained whose value mainly depends on the metal thickness. We have investigated the most simplest configurations using a homemade code based on the finite difference time domain method to design a half-plate with optimized efficiency resulting in transmission coefficient of more than $60\%$ together with a birefringence of $2.1$ at an operation wavelength of $737$ nm. This kind of anisotropic metasurface promises a wide range of applications in integrated photonics.
\end{abstract}
\maketitle
\section{Introduction}

Since the Ebbesen's experimental observation of the extraordinary transmission through a grating of nanoholes engraved into a silver film \cite{Ebbesen:nature98}, periodic sub-wavelength aperture structures have been proposed for different types of applications ranging from simple spectral filtering \cite{laux:np08,si:apl11} to the enhancement of spontaneous emission \cite{rigneault:prl05,aouani:nl11}, including the design of surface optical elements (lenses, wave plates, etc.) \cite{lezec:science02,li:apl08,baida:prb11,helgert:nl11}. In each application, the challenge consists in confining, enhancing or controlling light within these nanoscale structures \cite{baida:oc02, baida:oc05, baida:prb06, sun:ol06, poujet:ol07, banzer:oex10, ndao:apb12}. 
In the case of anisotropic metamaterials, the birefringence is artificially provoked through the geometry of the aperture that allows obtaining a phase difference between the two transverse components of the transmitted electric field. This results in compact ultrathin wave plates compatible with nano-optical devices, while exhibiting better performance than those made from natural birefringent materials especially in the terahertz domain such as low transmission, wide thickness and narrow bandwidth, making them unsuitable in miniaturized optical systems \cite{cescato:ao90,masson:ol06,chen:oc13}.\\
As mentioned before, the most efficient wave plates were obtained by breaking the symmetry of structures exhibiting extraordinary transmission (aperture in opaque films for instance) in the transverse plane \cite{gordon:prl04, gordon:prb06, baida:prb11,boutria:prb12,zahia:pb18,khoo:ol11}. In fact, pure plasmonic structures \cite{drezet:prl08, marcet:prl11, chen:oc13a,dai:oex14} have shown modest performances due to the weak transmitted light intensity that accompanies the plasmon excitation (inherent dissipation effect). Further studies achieved this anisotropy through guided mode excitation \cite{baida:prb11,dahdah:ieee12,wang:ol14,hu:oex17}, where the dissipative effects are minimized.\\
Unfortunately, two-dimensional metamaterials were widely proposed to design wave plates \cite{yang:plas11,wang:ol14,ding:acs15,owiti:rsc17,mustafa:sr18} in the near infrared range and THz frequencies \cite{li:oex17,sieber:oex14}, while very few studies have been devoted to the visible spectral range \cite{dahdah:ieee12}. In the later paper, a quarter-wave plate (QWP) based on an array of coaxial apertures with an elliptical core is proposed to operate at $\lambda=544$ nm. Nevertheless, the design of a compact half-wave plate in the visible range, characterized by a phase difference ($\Delta\phi=\pi$) between the two transverse electrical components at the output of the plate, is extremely difficult with such a configuration \cite{dahdah:ieee12}.\\

Based on enhanced transmission metamaterial as described above, we propose a simple design where the geometrical and physical parameters are optimized in order to achieve an operating half-wave plate in the visible range. In order to increase the value of the output phase difference $\Delta\phi$, one of two apertures has been filled by a dielectric (glass).\\ 
To the best of our knowledge, this is the first time that such simple-geometry half-wave plate (HWP) based on enhanced-transmission metamaterial is designed and optimized to operate in the visible spectral range (operation wavelength of $\lambda=737$ nm, bandwidth of $\Delta\lambda=50$ nm). Its transmission coefficient reaches 60\% while its thickness does not exceed $\lambda/4.2$. This will contribute to the development of basic optical components to be compatible with integrated photonics.

\section{Design and principle}

\begin{figure}[h!]
\centering
\includegraphics[width=.5\linewidth]{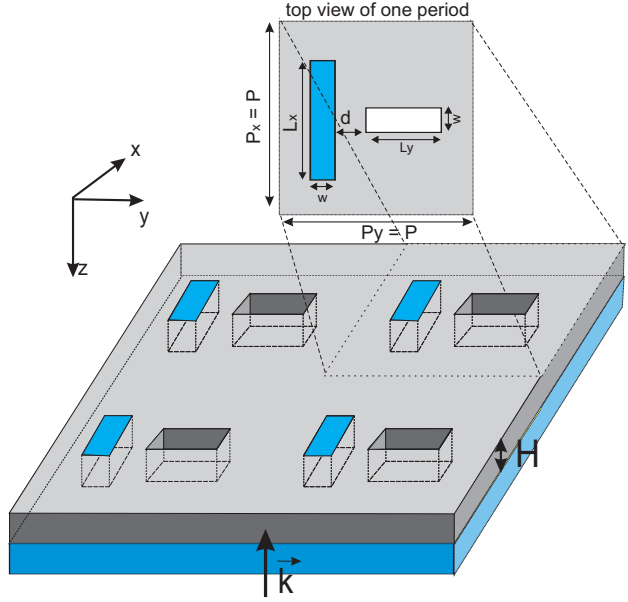}
\caption{Schematic of the proposed plate which consists of a bi-periodic grating with two perpendicular rectangular apertures per period engraved into an Ag film (gray) of thickness $H$ and deposited on a glass substrate (blue). A top view of one period is given at the top of the figure giving all the structural parameters.}\label{Figure1}
\end{figure}

The proposed configuration is depicted in figure \ref{Figure1}: it is composed of a 2D subwavelength double-pattern rectangular aperture array etched into an opaque silver (Ag) film deposited on a glass substrate (refractive index $n=1.5$). $H$ and $P$ denote respectively the Ag layer thickness and the structure period along both $x$ and $y$ directions. The array pattern consists of two perpendicular rectangular apertures. One of them is parallel to $x$-axis and is filled with glass of the same refractive index as the substrate. Its geometry is defined by its length $L_{x}$ and its width $W$. The second rectangular aperture is parallel to $y$-axis and defined by its length $L_{y}$ and has the same width $W$ as indicated in top of figure \ref{Figure1}. The distance $d$, along the $y$-axis, between these two cavities was fixed to $d=40$ nm.\\
Numerical simulations are performed using home-made finite difference time domain (FDTD) code, where periodic boundary conditions are applied in the $x$ and $y$ directions and perfectly matched layers (PML) are used in the $z$ direction as absorbing boundary conditions. A linearly polarized plane wave is normally incident from the substrate side. Its polarization direction is
identiﬁed with respect to the x-axis.\\

As well known, when illuminated by an electromagnetic wave, the apertures behave as waveguides allowing the excitation and the propagation of guided modes along the metal thickness under some conditions \cite{baida:prb06}. Indeed, the fundamental guided mode, named as $TE_{10}$ mode, can only be excited by the electric field component parallel to the width of the rectangular section. Its cutoff wavelength, in the case of a perfect conductor, is equal to twice its rectangular section length (here $L_x$ or $L_y$). In the case of metal with absorption (real metal), the cutoff wavelength value increases as its effective length due to the small, but not negligible, penetration of the electromagnetic field inside the metal. When excited at the cutoff wavelength, this guided mode will propagate along the metal thickness with a very small effective index (both real and imaginary parts). Its phase velocity tends to infinity meaning a constant value of its phase during the propagation. Consequently, the phase matching condition between two round trips of the light is automatically fulfilled leading to a maximum transmission as in Fabry-Perot interferometer. Having said that, the transmission spectrum of one rectangular aperture array will exhibit a peak at the cutoff wavelength provided that its electric field has a non-zero component along the rectangle width direction (noted by $W$). The transmission efficiency at this peak depends on the imaginary part of the guided mode effective index and on the metal thickness. Thus, when the thickness of the metal film increases, the transmission decreases but, more interestingly, other transmission peaks will appear in the spectrum (Fabry-Perot harmonics) verifying a phase matching condition for which the effective index is no longer zero (see equation (8) in \cite{baida:prb11}). The phase difference, between the transmitted light at the cutoff wavelength ($m=0$) and at the first Fabry-Perot harmonic ($m=1$), is at the origin of our design. In order to get an HWP, this phase difference must be equal to $\pi$ and the two excited modes should correspond to two orthogonal polarization directions (the two transverse components of the transmitted field). To summarize, obtaining a half-wave requires fulfilling three conditions: (1) a guided mode at the cutoff for a given polarization (first rectangle), (2) a guided mode transmitted at the same wavelength for the first harmonic of the orthogonal polarization (second rectangle) and (3) a high and identical transmission efficiency (smallest metal thickness) for both modes.

This last condition will be all the more efficient as the transmitted energy is directed in a single direction. Thus, and due to the phenomenon of diffraction, one will always consider sub-wavelength structures (an operation wavelength greater than the period of the structure) leading to a single diffracted order parallel to the incident direction (diffracted zero order). The main challenge is then to design HWP operating in the visible range because this small value of the operation wavelength implies small dimensions of the apertures, which makes their manufacture very difficult. A compromise must be found between the thickness of the metal layer and the size of the apertures so as to be able to integrate them both in the same period of the structure. To this end, we propose to fill one of the aperture by a dielectric which will correspond optically (not geometrically) to increase its dimensions. This technique was used by Khoo et al. \cite{khoo:ol11}, which simultaneously excited the fundamental modes of the filled and unfilled cavities to obtain a quarter-wave plate in the visible range. Indeed, as it was mentioned above, the effective index tends to zero near the cutoff. It is therefore difficult to further increase the phase shift at the output of the structure, to reach $\pi$ when only the excitation of the fundamental modes are involved as in \cite{khoo:ol11}. Consequently, we propose to break this lock using a mode excited at its cutoff ($ m = 0 $) with the same mode but excited at the wavelength of its first harmonic ($m=1$).

\begin{figure}[h!]
\centering
\includegraphics[width=.7\linewidth]{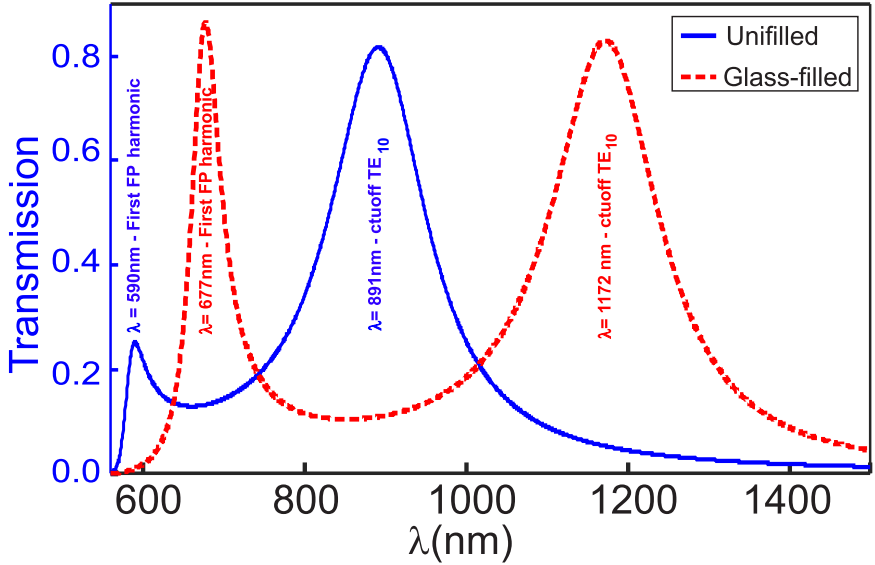}
\caption{Transmission spectra of an unfilled (blue solid line) and filled (red dashed line) rectangular cavities elongated along the x-axis when the incident plane wave is y-polarized. The geometrical paramters are: $P=350$ nm, $L_{x}=260$ nm, $W=60$ nm and $H =150$ nm. In both cases, the substrate is glass with $n=1.5$.}\label{Figure2}
\end{figure}

Our design process is organized through three steps: (a) design of the first aperture so that its cutoff wavelength corresponds to the operation wavelength, (b) design of the second aperture to exhibit a Fabry-Perot first-harmonic transmission peak at the same operation wavelength and (c) optimization of the double-aperture array to get a minimum coupling between the two apertures in order not to modify their transmission properties. 
Let us begin with the first step where only a cavity parallel to the x-axis and illuminated by a linearly polarized wave in the y-direction is considered. Two different situations are examined: glass filled ($n=1.5$) and unfilled (air filled) apertures. Figure \ref{Figure2} shows the calculated transmission spectra where the blue solid line corresponds to the air-filled aperture and the red dashed line for the glass-filled aperture.

\begin{figure}[h!]
\centering
\includegraphics[width=0.5\linewidth]{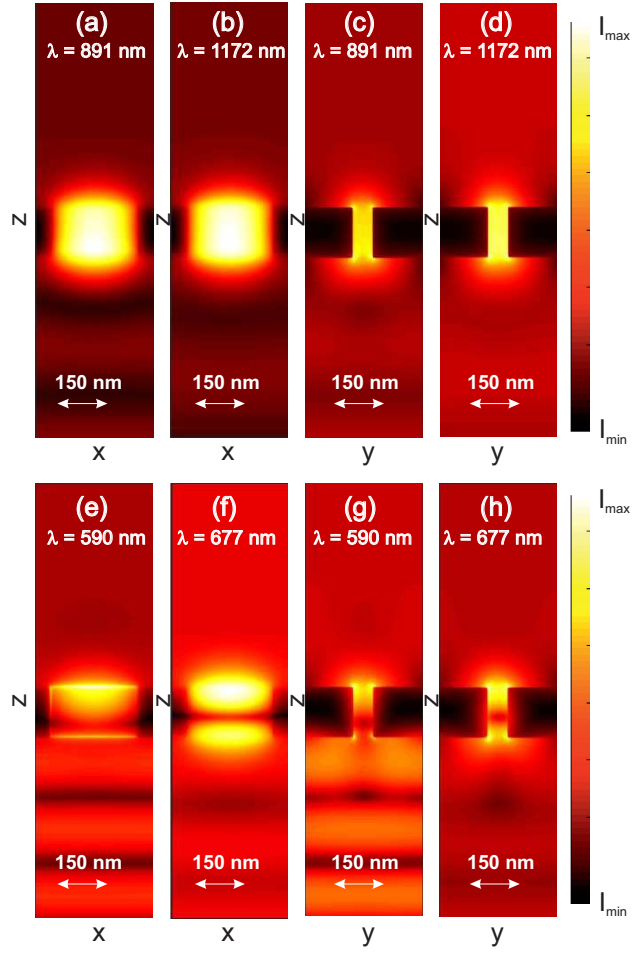}
\caption{Electric intensity distributions in vertical $xz$ (a,b,e,f) and $yz$ (c,d,g,h) planes passing by the center of the aperture. The wavelength value is set to: $\lambda=891$ nm in (a,c), $\lambda=1172$ nm in (b,d), $\lambda=590$ nm in (e,g) and $\lambda=677$ nm in (f,h). The structure has the same geometrical parameter as in figure \ref{Figure2} and it is illuminated at normal incidence by a y-polarized plane wave from the substrate.}\label{Figure3(a)_(h)}
\end{figure}

In order to verify the origin of these different peaks, we have mapped the distribution (in the x-z and y-z planes) of the electric field intensity at the four wavelengths corresponding to these peaks (see figure \ref{Figure3(a)_(h)}). Let us recall that, when the fundamental TE$_{10} $ mode is excited at its cutoff wavelength, the electromagnetic field phase remains constant during its propagation so that the round trips of light inside the aperture give rise to a uniform illumination \cite{baida:prb11}. As we can see, the mappings associated with wavelengths $891$ nm and $1172$ nm (see figure \ref{Figure3(a)_(h)}(a)-(d)) clearly correspond to the TE$_{10} $ fundamental mode excited at its cutoff wavelength ($m=0$), while the presence of intensity node inside the aperture for the wavelengths $590$ nm and $677$ nm (see figure \ref{Figure3(a)_(h)}(e)-(h)) indicates the excitation of a Fabry-Perot (FP) first harmonic ($m=1$) of the same TE$_{10}$ fundamental mode. 

The next step is to design a second aperture for which the first FP harmonic ($m=1$) of the guided mode spectrally matches the cutoff wavelength ($m=0$) of the first aperture TE$_{10}$ mode. This can be done easily because the cutoff wavelength is practically insensitive to the metal thickness (only depends on the aperture length value), whereas the spectral position of the FP harmonics ($m>1$)  strongly depends on $H$ through a conventional FP phase matching condition that explicitly involves $H$. Therefore, by adjusting the latter, the first FP harmonic ($m=1$) of the fundamental mode of the glass-filled aperture and the fundamental mode ($m=0$) of the air-filled aperture may overlap. Only by combining such two apertures filled by different materials (here air and glass), a large effective index difference could be obtained providing a phase difference of $PD=\pi$.\\

\begin{figure}[h!]
\centering
\includegraphics[width=0.7\linewidth]{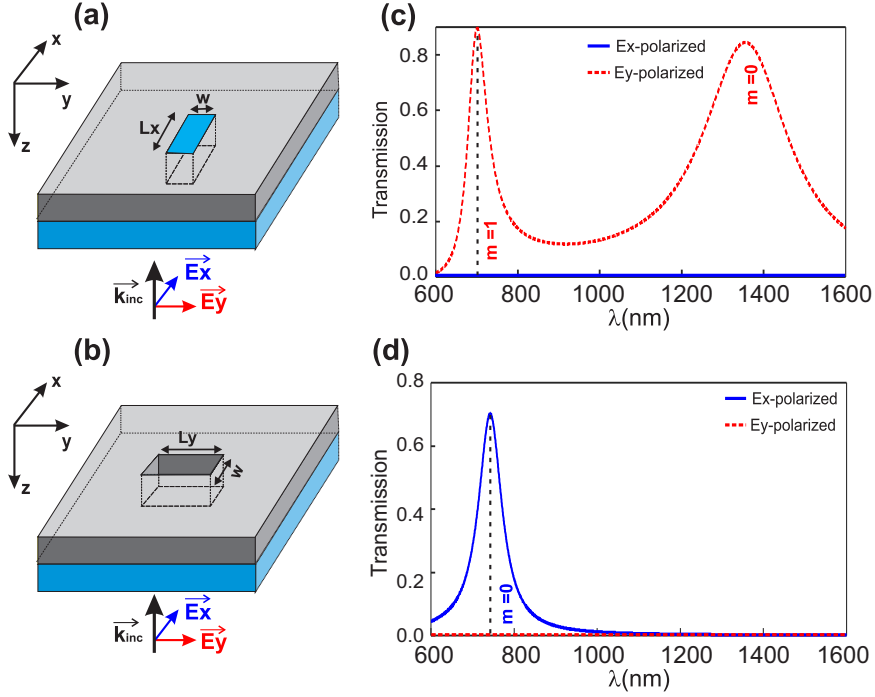}
\caption{Schematics of a RAA parallel to x-axis (a) and parallel to y-axis (b) engraved into a $H$-thick silver layer deposited on glass substrate. Geometrical parameters are: $ P =350$ nm, $ L_{x}=300$ nm, $ L_{y} =200$ nm, $ W =60$ nm and $H=150$ nm. (c) and (d) are the transmission spectra through (a) and (b) gratings when successively illuminated by linearly polarized plane wave along y- and x- axes, respectively. Two FP harmonics ($m=0$ and $m=1$) are excited in (c) while only the mode at the cutoff wavelength ($m=0$) is present in (d).}\label{Figure4}
\end{figure}

Figure \ref{Figure4} illustrates the excitation of the orthogonally polarized guided modes in the two apertures when they are considered separately: in (a) the grating is composed of only the glass-filled aperture with $L_x=300$ nm, $W=60$ nm, $H=150$ nm and $P=350$ nm. In (b), the grating pattern consists of an orthogonal aperture with $L_y=200$ nm, and the same values of $W$, $H$ and $P$. The transmission spectra presented on figures \ref{Figure4}(c) and (d) clearly show the occurrence of peaks when the incident polarization is directed toward the width (small side) of the aperture. More importantly, one can see on figure \ref{Figure4}(c) the excitation of the first FP harmonic of the TE$_{10}$ mode of the glass-filled aperture (see vertical dashed black line) in the visible range while the same peak of the air-filled aperture is out from the considered spectral range because it appears at smaller value of the wavelength. Only the peak corresponding to the excitation of the guided mode at its cutoff exists (dashed vertical line in figure \ref{Figure4}(d)). When excited simultaneously at the same operation wavelength, these two modes propagate with different phase velocities inducing a phase difference $PD$ between the two orthogonal components of the electric field that could fulfill the one of a half-wave plate ($PD=\pi$). To this end, we need to vary the metal thickness $H$ until these two peaks coincide while monitoring the value of $PD$.\\

\begin{figure}[h!]
\centering
\includegraphics[width=.9\linewidth]{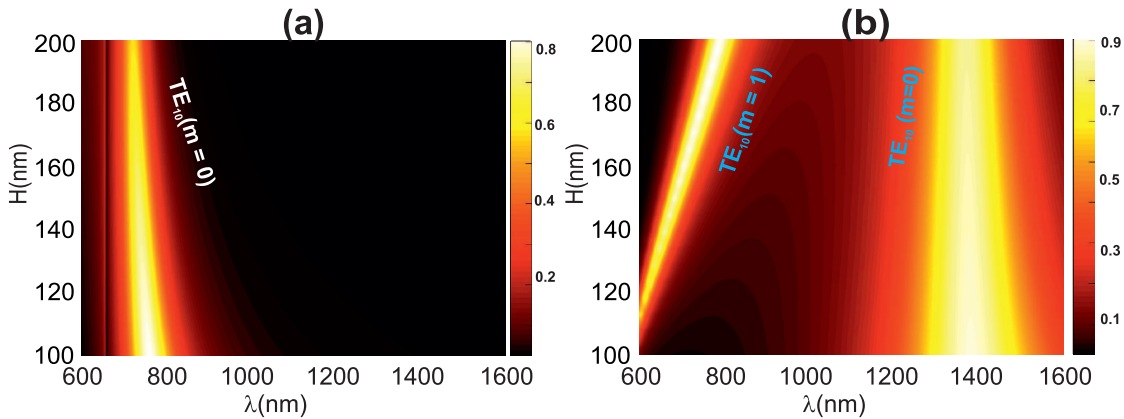}
\caption{Transmission spectra through the structure defined in figure \ref{Figure1} with $P=350$ nm, $L_{x}=300$ nm, $L_{y}=200$ nm and $W=60$ nm. The incident plane wave is x-polarized in (a) and y-polarized in (b).}\label{Figure5}
\end{figure}

Consequently, we performed numerical simulations and calculate the transmission spectra through the whole (2 apertures) structure defined in figure \ref{Figure1} by varying the metal film thickness for the two  polarization directions ($Ox$ and $Oy$) of the incident wave (see figure \ref{Figure5}).\\

 As expected, one can clearly see from figure \ref{Figure5}(b) that only the first FP harmonic ($m=1$) peak is greatly affected by the metal thickness value. The spectral position of the two peaks at the guided mode cutoff wavelength ($m=0$) remain quite constant. This sensitivity of the FP harmonic to the variation of the metal thickness will be exploited to adjust the phase value.

\section{Results and discussions}
From figure \ref{Figure5}, one can see that for metal layer with thickness between $H=160$ nm and $H=200$ nm, the two peaks associated with the first FP harmonic ($m=1$) of the rectangular aperture parallel to the x-axis and the TE$_{10} $ fundamental mode ($m=0$) of the rectangular aperture parallel to the y-axis have a significant overlap around $\lambda=700$ nm. A more precise and quantitative analysis of this overlap area will allow the determination of the thickness value providing the desired phase difference of $PD=\pi$.\\

Consequently, we will express the transmission properties of the whole structure through the transmission Jones matrix $t$ defined by:

\begin{eqnarray}
\vec{E}_{t}=
\left( \begin{array}{c}
E_{tx}  \\
E_{ty} \end{array} \right)
=
t \vec{E}_{inc}=
 \left( \begin{array}{cc}
\vert t_{xx} \vert e^{i\phi_{xx}} & \vert t_{xy} \vert e^{i\phi_{xy}} \\
\vert t_{yx} \vert e^{i\phi_{yx}} & \vert t_{yy} \vert e^{i\phi_{yy}}  \end{array} \right)
\left( \begin{array}{c}
E_{inc x}  \\
E_{inc y} \end{array} \right)
\end{eqnarray}
\newline

where $\vec{E}_{t}$ is the transmitted electric field,  $\vec{E}_{inc}$ is the incident electric field, $t_{ij}$ (where $(i,j)=(x,y)$) is the complex transmission coefficient defined as the ratio of the amplitude of the transmitted $j$ component of the electric field to the same quantity of the incident field that is polarized along the $i$ direction. Accordingly, $t_{ij}$ with $i\neq j$ (non-diagonal elements) correspond to the depolarization terms that should be very small compared to the diagonal ones as demonstrated in figure \ref{Figure4}. The axis-symmetry of the structure ($'T'$ shape) is at the origin of the weak coupling between the two apertures \cite{boyer:josaa14} and allows a simplified optimization of the structure geometry. In this case, the phase difference between the two transmitted transverse components of the electric field becomes $PD=\pm(\phi_{yy}-\phi_{xx})$. We recall here that the transmission Jones matrix associated to a perfect HWP is given by:
\newline
\begin{eqnarray}
t_{\lambda/2}=
 \left( \begin{array}{cc}
t_{xx} & t_{xy} \\
t_{yx} & t_{yy}  \end{array} \right) =
\left( \begin{array}{cc}
1 & 0 \\
0 & -1  \end{array} \right)\label{jones}
\end{eqnarray}

\begin{figure}[h!]
\centering
\includegraphics[width=0.7\linewidth]{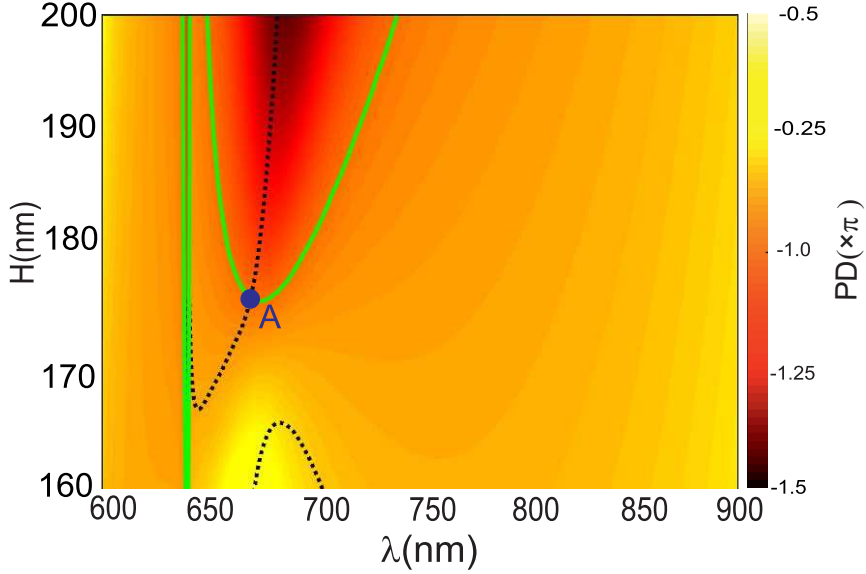}
\caption{Diagram of phase difference $PD= \phi_{yy}-\phi_{xx} $ as a function of the metal thickness $H$ and the wavelength $\lambda $. The solid green line corresponds to $DP=-\pi$ and the dotted black line corresponds to the couples ($\lambda,H$) that fulfill $ \vert t_{xx} \vert = \vert t_{yy} \vert$. The intersection point A corresponds then to the case of HWP. The geometrical paramters of the structure are $P=350$ nm, $L_{x}=300$ nm, $ L_{y}=200$ nm and $W=60$ nm.}\label{Figure6}
\end{figure}

To determine the four elements of $t$, two different numerical simulations per $H$ value are performed corresponding to the eigen polarization directions $Ox$ and $Oy$ ($j=x$ then $j=y$ in equation (\ref{jones})). Within the obtained data, we have all needed information to optimize our structure. Let us recall the conditions to get a half-wave plate (HWP): a large and equal transmission coefficients for both $x$ and $y$ polarization directions together with a phase difference of $\pi$. These two conditions can be formally stated by: 
\newline
\begin{eqnarray}
\vert t_{xx} \vert=\vert t_{yy} \vert  \hspace{1cm}  \mbox{and}  \hspace{1cm}  \phi_{yy}-\phi_{xx}=\pm\pi \label{conditions}
\end{eqnarray}
\newline
The numerical results are presented on figure \ref{Figure6} as a diagram giving the $PD$ value as a function of $\lambda$ and $H$. The dotted black line denotes the couples ($\lambda, H$) that fulfill the first condition of equation (\ref{conditions}) while the solid green line addresses the second condition of the same equation ($DP=-\pi$). The A point (blue circle), that is the intersection of these two curves, corresponds then to the case of HWP. From figure \ref{Figure6}, we get $A(H_A=175$ nm; $\lambda_A=737$ nm$)$ for which the Jones matrix $t_A$ is equal to:
\newline
\begin{eqnarray}
t_A=t(H_A=175\hspace{0.2cm}\mbox{nm},\lambda_A=737\hspace{0.2cm}\mbox{nm})=
 \left( \begin{array}{cc}
0.779 & 0.001e^{i0.587\pi} \\
0.003e^{i0.78\pi} & 0.782e^{-i0.98\pi}  \end{array} \right)\label{pointA}
\end{eqnarray}
\newline

As expected, the two diagonal terms have almost the same amplitude of $0.78$ and a phase difference of $PD=-0.98\pi$. The non-diagonal elements $ \vert t_{yx}\vert$ and $ \vert t_{xy}\vert$ are very small compared to the diagonal terms. This reflects a very weak coupling between the apertures and makes the proposed plate closer to a perfect HWP. The associated birefringence is estimated through its classical expression $\Delta n=\frac{\lambda_A PD}{2\pi H_A}$ to be 2.1 corresponding to an extraordinary value compared to natural birefringence ($\Delta n_{quartz}<1\times 10^{-2}$ in the visible range for instance).\\

\begin{figure}[h!]
\centering
\includegraphics[width=.8\linewidth]{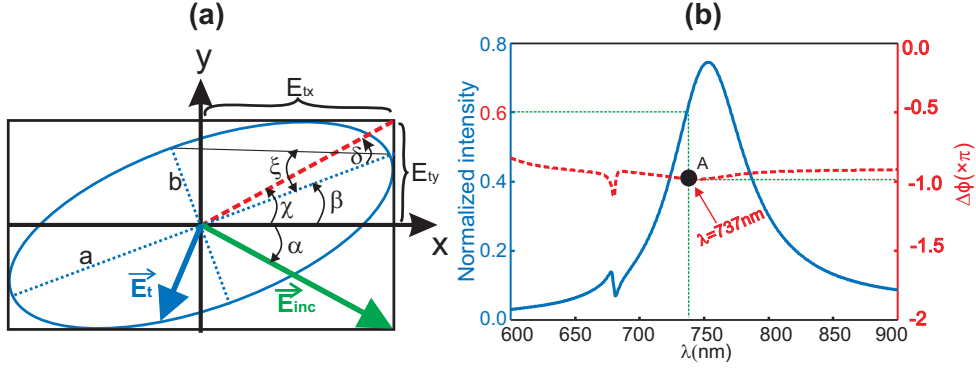}
\caption{ (a) Schematic of the polarization parameters of the incident ($\vec{E}_{inc}$) and the transmitted ($\vec{E}_{t}$) electric fields. The angles $\chi$ and $\zeta$ are directly linked to the transmission Jones matrix elements given by equation (\ref{jones}). Their expressions are provided by equations (\ref{angles}). (b) Transmission spectrum (solid blue curve) and phase difference (dashed red curve) in the case of $\alpha=45^\circ$ nm. The geometrical parameters are: $P=350$ nm, $L_{x}=300$ nm, $ L_{y}=200$ nm, $W=60$ nm and $H=175$ nm.}\label{Figure7}
\end{figure}

When an incident wave is linearly polarized toward an angle $\alpha$ counted from a perfect HWP axes (defined by the direction of the rectangular apertures i.e. $Ox$ and $Oy$ directions here), the transmitted wave polarization is rotated by an angle of $2\alpha$ and still is linear. For a real HWP, the transmitted wave is elliptically polarized. Consequently, the deviation between a perfect and a real HWP can be characterized through two quantities: the ellipticity $\eta$ of the transmitted wave polarization and the error on the ellipse axis direction $\delta=|\beta+\alpha|$ (see figure \ref{Figure7}(a)). These two quantities that are usually used to quantify the HWP properties \cite{baida:prb11}, are given by:

\begin{eqnarray}
\centering
\eta=tan(\zeta)=\frac{b}{a} \hspace{1cm} \mbox{and} \hspace{1cm} tan(2\beta)=tan(2\chi)cos(\Delta\phi)\label{angles}
 \end{eqnarray}
 \newline
where $a$ and $b$ are respectively the half major- and minor-axis values of the ellipse, $\beta$ is the angle between the ellipse major-axis and the HWP axis ($Ox$ here), $\Delta\phi=\phi_{yy}-\phi_{xx}$ is the phase difference between the $x-$ and $y-$transmitted components of the electric field and $\chi$ is given by $\tan(\chi)=\frac{|E_{ty}|}{|E_{tx}|}$. $\eta$ is calculated through the value of $\zeta$ that can be expressed as a function of $\chi$ and $\Delta\phi$ by $\sin(2\zeta)=\sin(2\chi)\sin(\Delta\phi)$. All these angles are depicted on figure \ref{Figure7}(a).\\

Numerically, we have fixed $\alpha$ to $45^\circ$ which corresponds to the case where the deviation is maximum. The spectral response of the transmitted wave in terms of normalized intensity and phase difference $\Delta\phi$ are shown in figure \ref{Figure7}(b) for the same structure with $H=H_A=175$ nm.\\

As it is shown in figure \ref{Figure7}(b) where are presented both the transmission spectrum and the phase difference for the same structure with $H=175$ nm, at the operation wavelength ($\lambda_A=737$ nm), the transmission coefficient is about $60\%$ and the $PD$ is equal to $-0.98\pi$. These results are in very good agreement with those obtained through the transmission Jones matrix value given by the equation (\ref{pointA}) demonstrating, once again, the very weak coupling between the two apertures even for $\alpha=45^\circ$ (the $x$ and $y$ incident electric field components are non zero). \\

\begin{figure}[h!]
\centering
\includegraphics[width=.95\linewidth]{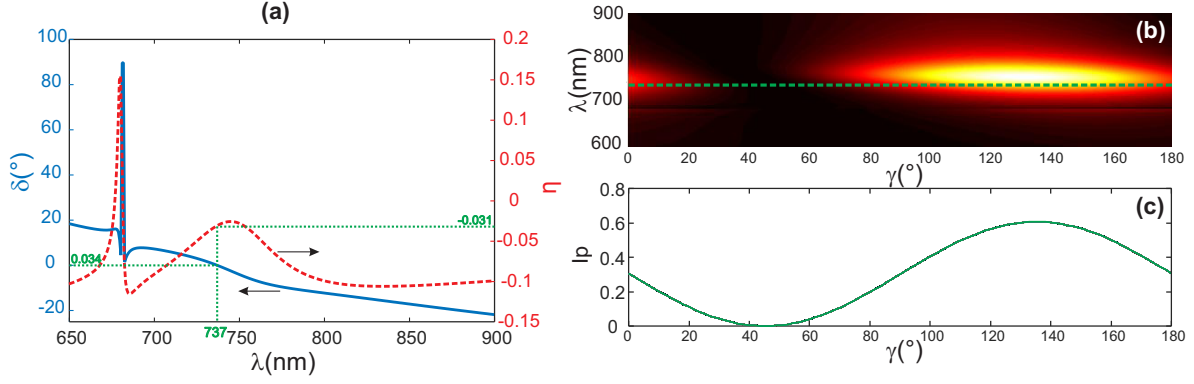}

\caption{ (a) Polarization properties of the transmitted field: the solid blue line gives the deviation $\delta$ of the rotation angle introduced by our structure with regard to the one of a perfect HWP while the red dashed line corresponds to the ellipticity $\eta$. (b) Map of the transmitted intensity $I_p$ through a polarizer placed at the output side of the structure with axis at an angle $\gamma$ with the $x-$axis. (c) Cross-section made on the map in (b) at the operation wavelength $\lambda_A$ demonstrating the good accuracy of the HWP. }\label{Figure8}
\end{figure}

In order to evaluate the properties of the designed HWP, figure \ref{Figure8} reports the properties of the transmitted wave in terms of polarization and efficiency (transmitted intensity). Figure \ref{Figure8}(a) gives the variations of the ellipticity and the deviation angle $\delta$ around the operation wavelength $\lambda_A=737$ nm. Figure \ref{Figure8}(b) presents a map of the transmitted intensity $I_p$ through a polarizer directed toward an angle $\gamma$ with respect to the $x-$axis for $\gamma$ varying from $0$ to $180^\circ$. $I_p$ can be simply expressed by $I_p=|E_{tx}\cos(\gamma)|^2+|E_{ty}\sin(\gamma)|^2$. Figure \ref{Figure8}(c) gives the variations of this intensity as a function of the polarizer direction at the operation wavelength $\lambda_A=737$ nm. As expected, for $\gamma=45^\circ$, $I_p$ vanishes meaning that the polarizer axis is orthogonal to the polarization direction of the transmitted field. The latter must be necessarily parallel to the second bisector at $\beta=-45^\circ$, which corresponds to a total rotation angle of the polarization of $2\alpha$. Note that the extinction ratio by the polarizer, given by $I_p^{min}/I_p^{max}$, is evaluated to be equal to $0.003$ that corresponds to an almost linear polarization for the transmitted wave.

Finally, figure \ref{Figure8}(a) shows that for wavelength range from $\lambda=710$ nm to $\lambda=760$ nm, the ellipticity varies between $\eta=-0.08$ and $\eta=-0.025$ while the deviation angle $\delta$ changes in the range of $\pm5^\circ$. This leads to an operation bandwidth of $\Delta\lambda=50$ nm and attributes to our structure the property of spectral filtering in association to behave as a HWP.

\section{Conclusion}
In summary, through FDTD simulations, we have successfully designed an anisotropic metamaterial with sub-wavelength double rectangular aperture array engraved into a silver (Ag) metallic film deposited on a glass substrate. The optimized structure behaves as an efficient HWP in the visible range (operation wavelength of $\lambda=737$ nm) with a bandwidth of $\Delta\lambda=50$ nm. This results in a giant birefringence of around $\Delta n=2.1$ together with an efficiency transmission of 60\%. This kind of anisotropic metamaterial promises a wide range of applications in integrated photonics.

\end{document}